\documentclass[aps,pre,floatfix,10pt]{revtex4}
\usepackage{graphicx}% Include Figure files
\usepackage{dcolumn}% Align table columns on decimal point
\usepackage{bm}% bold math
\usepackage{latexsym}
\usepackage{amsmath}
\usepackage{amssymb}
\usepackage{color}
\usepackage[normalem]{ulem}
\usepackage{multirow}
\begin{document}
\title{{\bf Flagellar length fluctuations: quantitative dependence on length control mechanism}}
\author{Swayamshree Patra} 
\affiliation{Department of Physics, Indian
  Institute of Technology Kanpur, 208016, India} 
\author{Debashish Chowdhury{\footnote{E-mail: debch@iitk.ac.in}}}
\affiliation{Department of Physics, Indian Institute of Technology
  Kanpur, 208016, India}

%%%%%%%%%%%%%%%%%%%%%
\maketitle
%%%%%%%%%%%%%%%%%%%%%

Organelles of optimum size are crucial for proper functioning of a living cell. The cell employs various mechanisms for actively sensing and controlling  the size of its organelles \cite{marshall15}. Recently Bauer et al have opened a new research frontier in the field of subcellular size control by shedding light on the noise and fluctuations of organelles of controlled size \cite{bauer21}. 
Taking eukaryotic flagellum as a model organelle, which is quite popular for such studies because of its linear geometry and dynamic nature, Bauer et al have analysed the nature of fluctuations of its length. Here we summarize the key questions and the fundamental importance of the recent developments. Although our attention is focussed here mainly on the experimental \cite{bauer21} and theoretical \cite{bauer21,fai20,patra20,patra20b,patra21,bressloff18} works on eukaryotic flagellum, the ideas are general and applicable to wide varieties of cell organelle.

\begin{figure*}
\includegraphics[width=1.0\textwidth]{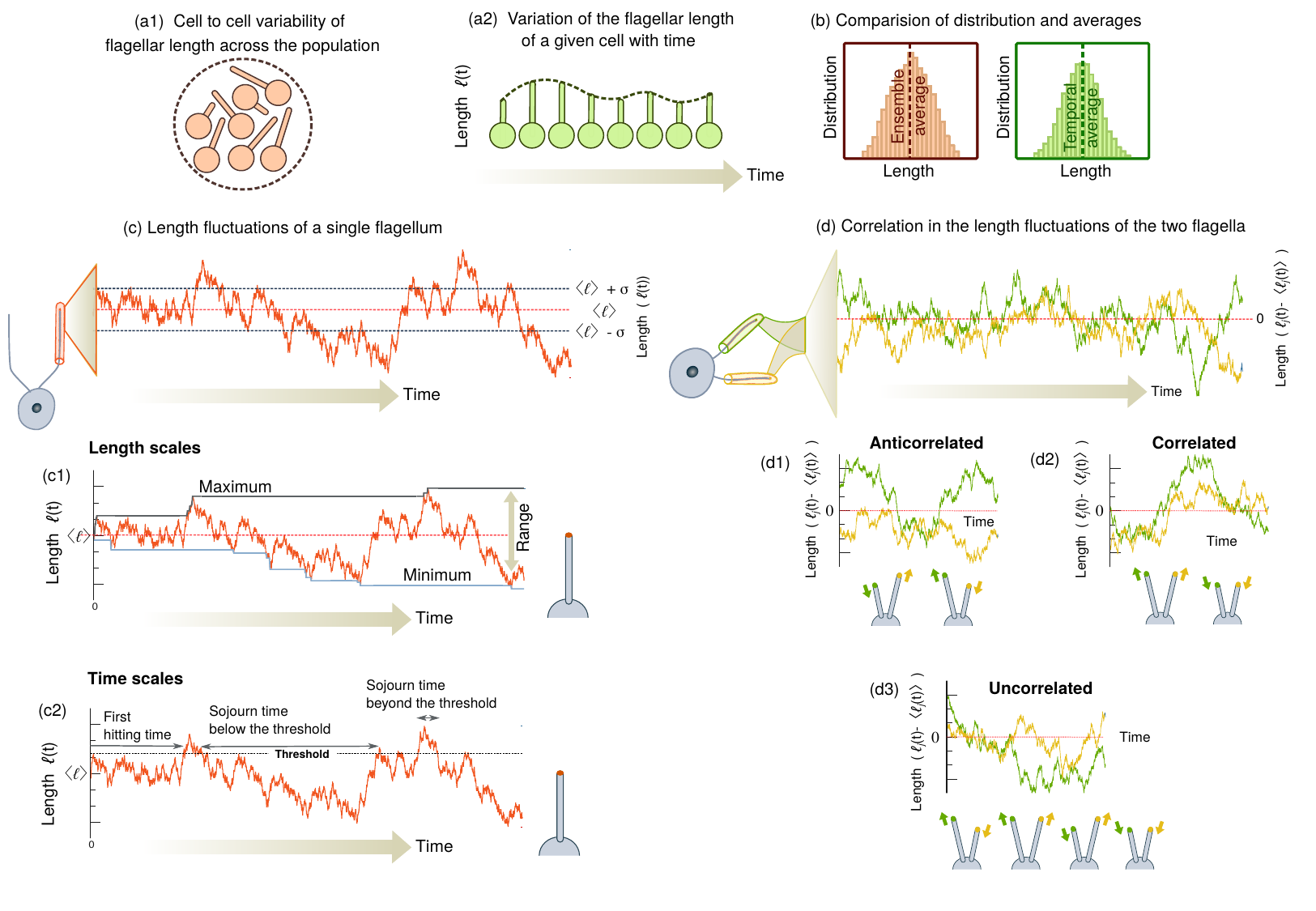}
\caption{{ {\bf Length fluctuations and correlations: }  }}
\label{fig1}
\end{figure*}

{\bf Controlled size: Ensemble average vs time average.} For the same species of unicellular organisms, or the same type of cells of a given multicellular species, the size of a  particular type of organelle can vary from one cell to another (Fig.\ref{fig1}(a1)). Such {\it cell-to-cell variations}, caused by noisy gene expressions and other factors, lead to the `heterogeneity' in the population of the same type of cells \cite{chang17}. Averaging over different copies of the organelles, or organelles from different cells give what is called the ``ensemble average'' in statistical physics. The size of a particular organelle in a given cell can also vary with time as well i.e, the organelle undergoes {\it temporal size fluctuations} (Fig.\ref{fig1}(a2)); averaging the size over time gives the ``time average''. In equilibrium statistical mechanics, systems for which time-average is identical to the corresponding ensemble-average are called {\it ergodic} (Fig.\ref{fig1}(b)). But, since living cells are far from thermodynamic equilibrium, there is no reason to expect the two averages to be identical. 

{\bf Why go beyond the average?}  In a population that is not too large none of the members of that ensemble may have a flagellum that is exactly as long as the {\it ensemble-averaged} flagellar length. Similarly, most of the time the instantaneous length of a flagellum may be slightly larger or smaller than the corresponding {\it time-averaged} value (Fig.\ref{fig1}(c)). Both these facts justify the need for looking  beyond the average. A beginning has been made in this direction through experimental studies by Bauer et al reported in their very recent paper \cite{bauer21}. 
Size fluctuations can be characterized in terms of several  statistical quantities which illuminate different aspects of the fluctuations and length control mechanism from different complementary perspectives \cite{patra20,fai20}.

{\bf Temporal size fluctuations of a single organelle.} Long cell protrusions often lack the machineries for the synthesis, degradation and recycling of its precursor proteins. Due to ongoing turnover of the building blocks  and probabilistic incorporation of fresh ones  at their tips, the length of  the protrusion fluctuates with the passage of time. The incessant random movement of the tip of the protrusion around its mean position is essentially a confined random walk. 

Mathematically speaking, taking the mean position of the protrusion tip as the origin, the random motion of the instantaneous position of the tip can be described as the Brownian motion of a hypothetical particle that is also  subjected to a restoring spring force (or, equivalently, a linear drift that tends to restore its position to the orgin).  
The incremental change $dL$ of the length in the infinitesimally small time interval $dt$ is best described by the  Ornstein-Uhlenbeck process
\begin{equation}
d L=-\frac{1}{\gamma}Ldt+\sqrt{2D}dW(t)
\end{equation}
where $\gamma$ is the relaxation time and $D$ the diffusion constant \cite{bauer21,patra20}.

%%%%%%%%%%%%%%%%%%%%%%%%%%%%%%%%
%%%%%%%%%%%%%%%%%%%%%%%%%%%%%%%%

Several competing models successfully predict the temporal evolution of the mean lengths of flagella. But, some of those models which predict the temporal evolution of the mean length correctly may have different predictions on the statistics of length fluctuations \cite{patra20,ludington15}. Therefore, the additional requirement of reproducing the experimentally observed statistics of fluctuations and noise would impose more stringent tests of the validity of the theoretical models. 

We list a few fundamental questions that need to be addressed.  For example, what are the maximum and the minimum length the organelle can attain over a given period of time (Fig.\ref{fig1}(c1))? What is the range it can scan during its lifetime (Fig.\ref{fig1}(c1))? Similarly, for a given threshold of interest, when does the tip of the protrusion hit the threshold for the first time (or, in general, $n$-th time) ((Fig.\ref{fig1}(c2))? For what fraction of its lifetime does its length lie above (or below) the threshold and what is the rate of upcrossing and downcrossing the threshold (Fig.\ref{fig1}(c2)). These questions are answered in terms of the statistical quantities \cite{patra20} that can be calculated by borrowing the techniques of level crossing statistics from the theory of stochastic processes. 
%Such techniques have been employed also for understanding the stochastic excursions of molecular motors on the %cytoskeletal filaments \cite{guillet20}.  
These statistical quantities  depend on the detailed mechanisms of length control assumed by the model. Therefore, comparison of these theoretical predictions with the corresponding experimental data would help in the elimination of at least some of the models of length control. 

{\bf Use of correlations for probing communications among the multiple organelles of a cell:} 
In a multi-flagellated cell the a flagellum may elongate during a short time interval during which another may shorten or vice-versa (see Fig.\ref{fig1}(d)). Correlation function, as the name suggests, is an indicator of the statistical (in-)dependence of the kinetics of lengths of the two flagella. The correlation functions also reveal which flagella share a common pool of precursors and how. Here we summarize some recent developments on both the theoretical and experimental fronts. 

(i) {\it Anticorrelated:} During the regeneration of a selectively amputated flagellum \cite{patra20b} or the replication of a daughter flagellum \cite{patra21}, the pool can get get severely depleted. The negative correlation observed \cite{patra20} indicates that, in such situations, one flagellum can elongate only if the other shortens simultaneously because the limited resources in the shared  pool of precursors is inadequate to support sustained growth of a flagellum (Fig.\ref{fig1}(d1)). 

(ii) {\it Correlated:} By the term `correlated' here we mean positively correlated. The interpretations of the counter-intuitive positive correlation (Fig.\ref{fig1}(d2))observed by Bauer et al \cite{bauer21} are, at present, speculative. For example, depolymerization of cytoplasmic microtubules could, in principle, supply tubulin to both the flagella for their growth resulting in the observed positive correlation. An alternative supply of large quantities tubulins, adequate for positive correlation, may come from translational bursting. The length fluctuations of the pair of daughter flagella undergoing ciliogenesis prior to  cell division in the monoflagellates, are also positively correlated \cite{patra21}.

(iii) {\it Uncorrelated:} Prior to the cell division, when the monoflagellate cells are in multiflagellated phase, the new daughter flagella emerge from a common pool shared by the existing mother flagellum as well. Depending upon the pair of flagella, initially the correlation in their length fluctuations can either be positive or negative. Slowly, the correlations die or the fluctuations become uncorrelated (Fig.\ref{fig1}(d3)) indicating gradual approach towards the complete separation of the precursor pools \cite{patra21}. 

In future, the measurement of  correlations in the fluctuations of lengths of distinct flagella can not only clarify how the anisokont octoflagellated cells like {\it Giardia} \cite{mcinally19} assemble and maintain their multiple flagella of unequal length but also reveal the rules of molecular communication among the flagella. \\

{\bf Acknowledgements}: This research is supported by SERB (India) through J.C. Bose National Fellowship of DC. \\

{\bf Author contributions}: SP and DC designed the research, SP created all the artwork in Fig.1, SP and DC wrote the paper.

\end{document}